# Atomic Structure of Hematite ($\alpha$-Fe$_2$O$_3$) Nanocube Surface; Synchrotron X-ray Diffraction Study

Chang-Yong Kim[1], *

[1] Canadian Light Source Inc., 44 Innovation Boulevard, Saskatoon, SK, Canada S7N 0X4

*Corresponding author. E-mail: Chang-Yong.Kim@lightsource.ca

## Abstract

Atomic structure of a mono-dispersive hematite ($\alpha$-Fe$_2$O$_3$) nanocube (01$\bar{1}$2) surface was determined with synchrotron X-ray diffraction. The $\alpha$-Fe$_2$O$_3$ nanocubes were prepared through a hydrothermal process and a single layer of nanocubes was deposited on a silicon substrate with a drop cast method. The alignment of nanocubes with their {01$\bar{1}$2} surfaces parallel to the substrate is confirmed with grazing incidence X-ray diffraction. Specular crystal truncation rods (CTR's) from as-prepared and vacuum annealed nanocube surfaces have been measured and they are drastically different from previously reported CTR's from macroscopic single crystal (01$\bar{1}$2) surfaces. The measured CTR's from nanocube surfaces are explained well with the atomic structure models of half of atoms in top Fe layer being missing while extra oxygen-layers cover the half-missing Fe layer. An acidic environment during hydrothermal nanocube synthesis process is proposed as the main cause of the difference.

# 1. Introduction

Metal-oxide nanoparticles have drawn great research interest due to their size-dependent property changes and wide applications in medical imaging [1], drug-delivery [2], and catalysis [3]. The crystallinity of metal-oxide nanoparticles has a significant effect on their properties and is often manifested through different reactivities depending on crystalline orientations of their surfaces [4-9]. Often, nanoparticles are used under aqueous, high-temperature or high-pressure gas environments. Occasionally nanoparticle surfaces are modified to cover toxic bare surfaces or to functionalize the nanoparticle [10, 11]. Although determining atomic structures or their changes under various environments or after functionalization is critical, the characterization of atomic structures of nanoparticle surfaces is difficult. Irregular nanoparticle shapes and the broad distribution of particle size are the main obstacles for the characterization of surface atomic structures. Recent determination of crystallographic faces of nanowire with a scanning tunneling microscope (STM) [12] and of surface termination with a scanning transmission electron microscope (STEM) [13] are rare examples of atomic-scale surface structure characterization. The lack of detail we have about the atomic structure of nanoparticles leads to the issue of material-size gap. This leads to the understanding that a single crystal model system study cannot be extended to explain a nanoparticle's surface behavior. Considering the size-dependent nanoparticle properties [14], monodispersive nanoparticles can potentially eliminate the complexity caused by broad size distribution in the characterization of various physical and chemical properties of nanoparticles. Uniform particle shapes confined by low index atomic planes provide the opportunity to study chemical reactions specific to a crystallographic surface orientation in the nanoparticle-scale.

Iron oxides are one of the most studied nanoparticles for morphology control [15-18] and their surface structures and reactivities have been studied extensively [19]. Hematite ($\alpha$-$Fe_2O_3$) is the most stable form of iron oxide and various monodispersive, uniform shape nanoparticles have been synthesized [7, 20-27]. For example, a hematite nanoplate has large hexagonal (0001) surfaces as its base planes [22, 23] and a nano-rhombohedra is enclosed by six $\{10\bar{1}4\}$ crystallographic surfaces [20]. In particular, a pseudo-cubic nanoparticle (hereafter called a nanocube) has all six faces with crystallographically identical $\{01\bar{1}2\}$ planes as shown in Fig. 1(a) [25-27]. A single layer of monodispersive nanocubes on a substrate contacts one of the $\{01\bar{1}2\}$ faces with the substrate giving a well-aligned top surface normal direction of nanocubes [9]. Then, surface X-ray diffraction [28, 29] from the nanocubes along the specular direction can be measured as the atomic structure and direction of the top surface of an individual nanocube are more or less identical. Since the main information comes from atoms at the near surface region (a small portion of bulk sample), the surface X-ray diffraction has typically been performed with a synchrotron X-ray source which provides a much higher X-ray flux density (X-ray intensity per unit area) compared to a laboratory X-ray source [30]. An advantage that X-ray diffraction holds over other surface characterization tools is that, due to its weak interaction with matter, it can investigate surface atomic structures in contact with ambient gas or aqueous environments or structural changes under high temperature and/or high pressure conditions. Previous X-ray diffraction studies from hematite $(01\bar{1}2)$ single crystal surfaces under hydrous or saturated humidity environment have reported sample preparation dependent surface terminations [31-34]. Comparing $(01\bar{1}2)$ the surface structures of nanocubes and a macroscopic single crystal $(01\bar{1}2)$ will give a clue to bridge materials size gap.

We used surface X-ray diffraction to study the surface atomic structure of hematite nanocubes in its hydrated and vacuum annealed phases.

## 2. Material and methods

Hematite nanocubes have been synthesized following the recipe in literature [25]. Briefly, 0.721 g of $Fe(NO_3)_3 \cdot 9H_2O$ and 1.071 g of poly(Nvinyl-2-pyrrolidone) (PVP, Mw=30000) were dissolved in 64.29 mL of N,N-dimethyformamide (DMF), then placed in a Teflon-lined stainless steel autoclave of volume 125 mL. The sealed autoclave was put into an oven and heated at 180 °C for 30 h, and then cooled to room temperature naturally. The powder was obtained by centrifuging the mixture and washing the sediment four times with water. The washed nanocubes were re-dispersed in deionized water to make a nanocube solution of red color. The substrate was prepared by annealing a Si(001) wafer at 700 °C under $O_2$ flow for 3 hours, dipping it into 90 °C boiling water for 1 hour, and washing it with 10 % HCl. After blown dry with ultra-high purity nitrogen, the substrate was dried at 200 °C under flow of nitrogen. The sample was prepared by placing drops of nanocube solution on the hydroxylated silicon wafer and dried naturally. The drop-dry process was repeated a few times to get a dark-red color film.

X-ray diffraction measurements were performed at the HXMA beamline in Canadian Light Source with a 10.0 keV X-ray selected with an Si (111) double crystal monochromator. The X-ray was focused to sample position with Pt-coated toroidal mirror. The beam size defined by the slits in front of the sample was 0.2 x 0.5 mm$^2$ (V x H). Integrated intensities for crystal truncation rods were obtained with a rocking scan of the sample mounted on a 6-circle diffractometer (Huber). Diffraction intensities passing through two sets of horizontal and vertical slits were measured with a NaI detector (cyberstar, Oxford). For grazing incidence X-ray diffraction, the incidence angle of the X-ray to sample surface was set to 1.0 degrees and diffraction images were collected with a X-ray CCD detector (SX165, Rayonix) located about 15 cm from the sample.

Vacuum annealing was performed inside a small X-ray scattering chamber with a hemispherical beryllium dome. The X-ray scattering chamber was equipped with a pyrolytic boron nitride heater and pumped by a turbo molecular pump to get pressure better than $1.0 \times 10^{-7}$ Torr. In-situ crystal truncation rod (CTR) measurement from annealed surface was carried out after the sample was cooled down to room temperature under the $1.0 \times 10^{-7}$ Torr vacuum.

Scanning electron microscope (SEM, Zeiss EVO) images were taken with a secondary electron detector and an electron acceleration voltage of 20 kV.

### 3. Results

The nanocubes are of a pseudo-cubic shape instead of a perfect cube as shown in Fig 1(a). Each face of a nanocube is a parallelogram shape with edge angles of 85.66 ° and 94.34 ° and faces in the opposite sides of a nanocube are parallel to each other. Fig. 1(b) shows the grazing incidence X-ray diffraction (GIXRD) pattern from the drop-casted nanocube layer, which indicates that nanocubes are layered with their (01$\bar{1}$2) faces touching the substrate. The nanocube size was estimated to be 48 nanometers based on the SEM image (in Fig. 1(b) the vertical lines marking opposite sides of a nanocube are separated by 48 nanometers). Since azimuthal orientations of nanocubes are random, the nanocube-layer can be considered as a two-dimensional (2D) powder with a well-defined surface normal direction. The crystallographic $hk\bar{\imath}l$ indexes have been assigned to representative Bragg spots. Nanocube orientation can be made random (3D powder) by an intentional stir of drop-casted droplet and a uniform ring pattern was observed as shown in Fig.1(c). Because nanocubes are not perfect cubes, $(1\bar{1}02)$ and $(\bar{1}102)$ diffraction peaks appear slightly above the horizon (Fig. 1(d)), which would be on the the horizon if nanocubes were of a perfect cubic shape.

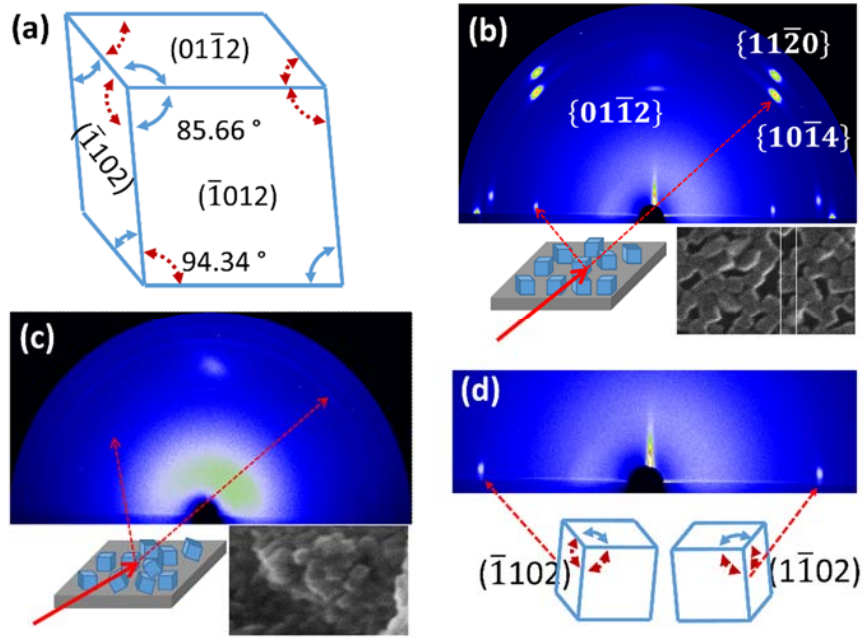

**Figure 1** (a) Shape of nanocube with indexes of crystallographic planes. Solid-blue and dotted-red arrows indicate edge angles of 85.66 ° and 94.34 °, respectively. Grazing incidence X-ray diffraction and SEM images from nanocubes (b) well-aligned along surface normal direction and (c) randomly oriented. Solid arrows indicate incident X-ray and dotted arrows represent diffracted X-ray. The lines in the SEM image in (b) are 48 nanometers apart. (d) Nanocubes orientations giving off-horizon $(1\bar{1}02)$ and $(\bar{1}102)$ diffraction peaks.

As shown in the SEM image in Fig. 1(b), a nanocube-layer can be seen as a highly textured uniform thickness film with voids exposing the bare substrate. The $(01\bar{1}2)$ Bragg peak from the nanocube layer shows a thickness fringe which manifest a narrow size distribution of the nanocube (Fig. 2). Since the $\{01\bar{1}2\}$ faces of nanocubes are aligned along the substrate normal direction, it is possible to measure CTR along the surface normal direction of nanocubes. Occasionally nanocubes sit on top of another nanocube or make a small cluster. Those voids and multilayered nanocubes make the surface rough. However, surface roughness affecting CTR is on the scale of several atomic-planes in length and typically modeled as stacking of progressively less occupied layers in the surface region (for example, layer occupation going 1/2, 1/4, and 1/16 as getting toward topmost surface layer) [35]. Also a CTR is results from coherent interference between atomic planes in surface region separated by an Angstrom length scale. In the case of multilayered nanocubes, the stacking of nanocubes would be far from a well aligned cube-on-cube arrangement and atomic planes in monolayer nanocubes and those in stacked-on nanocube are not aligned in Angstrom scale to induce the coherent interference. Such misaligned nanocubes would not contribute to CTR. If there is a good cube-on-cube arrangement, the top nanocube would contribute to CTR as the nanocubes in the sub-monolayer contribute. Multilayered nanocubes could cause a shadow effect to decrease overall CTR intensity. Overall, voids or multilayered nanocubes at the scale of tens of nanometers in length would reduce overall intensity without affecting CTR profile.

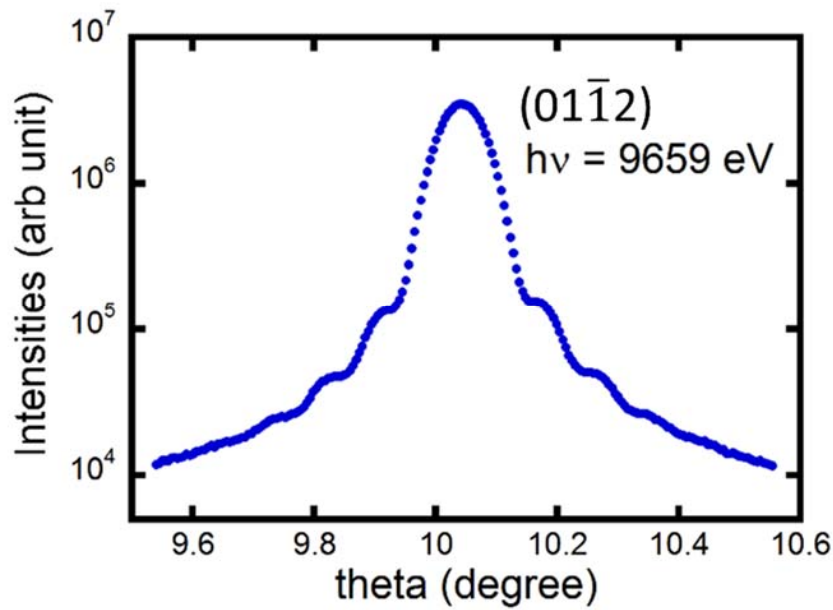

**Figure 2** X-ray diffraction around $(01\bar{1}2)$ peak from nanocube submonolayer.

To get the CTR profile, integrated intensities from sample rocking scans were measured at each q-point in the rod. The rocking curves were fit with a Voigt function with quadruple backgrounds and the Lorenz corrections applied to integrated intensities to get CTR profiles. The full widths at half maximum of the peaks from rocking scans were typically about 1.5 degrees. Peaks originated from the substrate CTR are much sharper in angular width and slight miscuts in the substrate gives slanted and split peaks. Hence, broad CTR peaks from the upper faces of nanocubes were easily separated from substrate peaks in the case that both CTR peaks from substrate and nanocube surface were detected during single rocking scan.

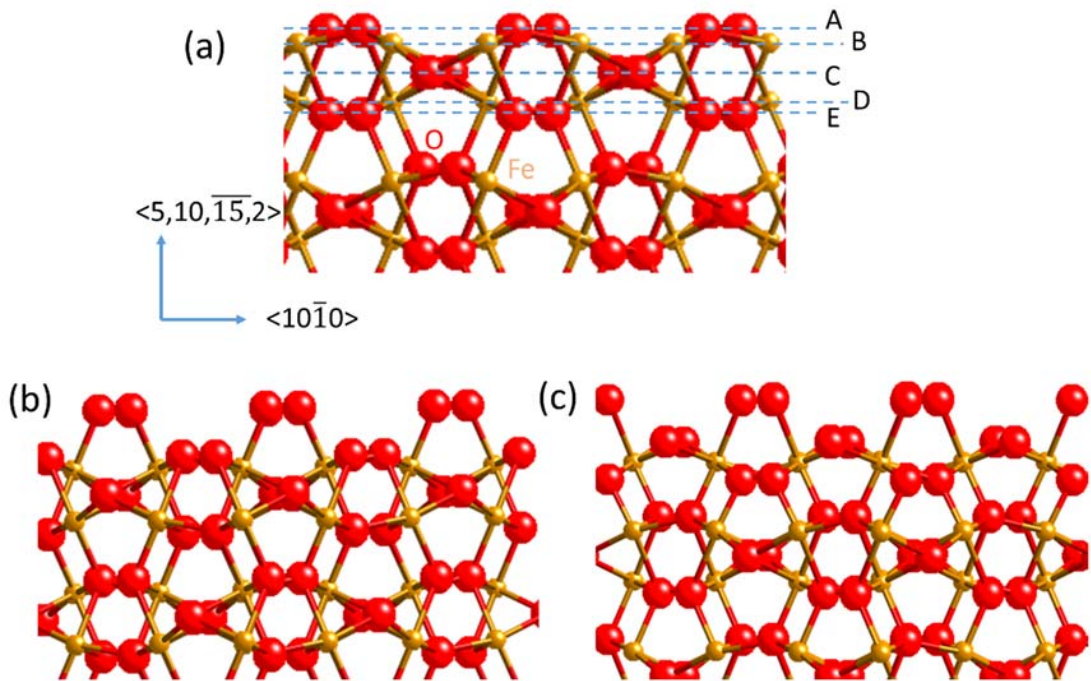

**Figure 3** Surface termination models. (a) Possible stoichiometric bulk terminations with dashed lines indicating topmost layers, (b) termination-E, and (c) termination-C with oxygen adlayer.

The $\alpha$-Fe$_2$O$_3$ (01$\bar{1}$2) surface can have five stoichiometric terminations as shown in Fig. 3(a) where the dashed lines indicate the topmost surface layer of each termination. In the literature, CTR measurements from single crystal $\alpha$-Fe$_2$O$_3$ (01$\bar{1}$2) surfaces under water and saturated humidity can be found [31-34]. From these measurements, three different surface terminations have been proposed. Ultra-high vacuum (UHV) prepared surface has termination-A and surface under saturated humidity has the termination-E shown in Fig. 3(b). Surface under water has water layers on top of the termination-E [33]. Chemomechanically polished surface was reported to have missing top Fe-layer (layer B) in the termination-A, which can be alternatively seen as the termination-C with oxygen adlayer [31, 34] as shown in Fig. 3(c). It was reported that annealing of the chemomechanically prepared sample in the air reverted surface to the termination-E [32].

Measured CTRs from as-prepared and vacuum annealed sample have been fit to the five stoichiometric terminations with an option of added oxygen layers including water ($H_2O$) and hydroxyl group (OH) adlayers. Since the nanocube layer was drop-casted from nanocubes dispersed in water and CTR from the as-prepared sample was measured under ambient condition, it would be safe to assume that the as-prepared surface is a hydrated surface. The as-prepared surface will be called a hydrated surface hereafter. The vacuum annealing temperature of 115 °C is not high enough to induce substantial structural change (e.g. converting from one type of termination to another). A model fit of hydrated and annealed surfaces was performed simultaneously. During the simultaneous fit, Fe occupations of same corresponding layers (for example B-layer in Fig. 3(a)) in both surfaces were maintained equal. However, occupations in top oxygen layers were allowed to differ from each other to consider dehydration by annealing (hydroxylation and desorption of water) of hydrated surface. Fits were performed using the GenX software which uses a genetic algorithm to avoid effectively being trapped in local minima [36, 37]. Goodness of fit was checked with the logarithmic R2 factor ($\sum_i (\log Y_i - \log S_i)^2 / \sum_i \log(Y_i)^2$, where $Y_i$ and $S_i$ are measured and simulated intensities respectively). By maintaining the layer-stacking order of the bulk during fit processes, models with atomic layers exchanged positions with adjacent layers were excluded.

No satisfactory fits were obtained with all the stoichiometric-termination models considered. Hence, non-stoichiometric terminations in which the occupation of top Fe layer (B-layer) was allowed to vary were considered. Again, Fe occupations were kept same for both hydrated and annealed surfaces. Fit was significantly improved with Fe occupation of ~50 % (half-Fe missing) as the logarithmic R2 factor reduced from typical value of 0.071 to 0.033. The best fits to the half-Fe missing model are shown in Fig. 4 along with a comparison with simulations based on reported models; the termination-C with oxygen adlayers [34] for hydrated and the termination-E [32] for annealed surface. Atomic layer distances obtained from fit to the half-Fe missing model are summarized in table 1 and 2.

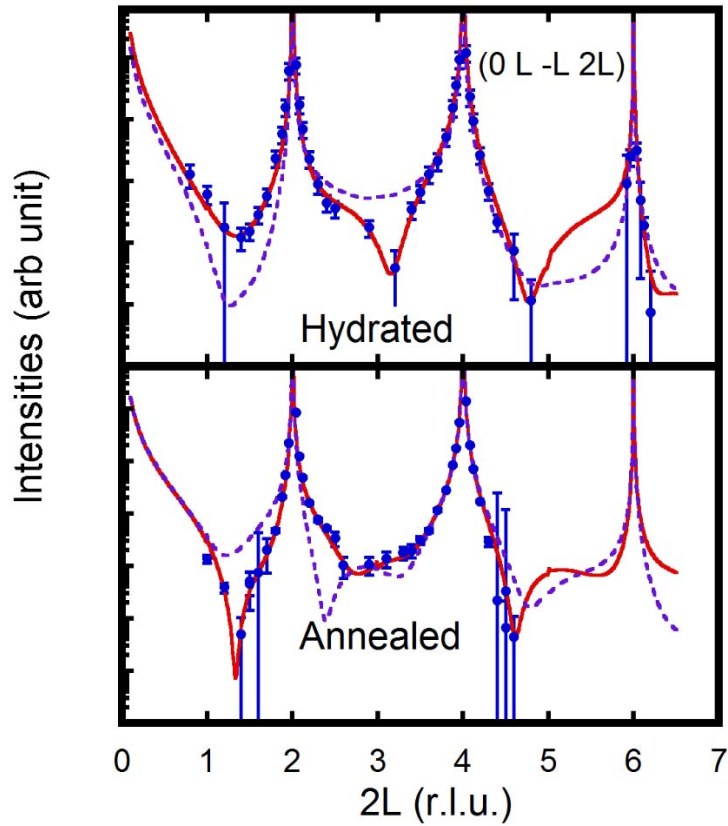

**Figure 4** Crystal truncation rods from nanocube submonolayer. Solid lines are best fit to half-Fe missing model (refer text for details). Dashed lines are simulations based on models reported by Ref. [34] (hydrated) and Ref. [32] (annealed).

**Table 1** Vertical positions and occupations of atomic layers and interlayer distances obtained from fit to the as-prepared surface CTR's. Vertical positions are in the unit of repeating vertical distance of 7.37 Å.

| Layer | Vertical positions and occupations | | | | Interlayer distances | | | |
|---|---|---|---|---|---|---|---|---|
| | Bulk | Nanocube | Changes | Occupations | Description | Bulk (Å) | Nanocube (Å) | Changes (%) |
| Ow | 1.750 | 1.69 (2) | -0.06 | 1.2(2) | Ow-O1 | 1.13 | 0.93 | -17 |
| O1 | 1.597 | 1.57 (1) | -0.03 | 0.7(2) | O1-O2 | 1.43 | 1.32 | -7 |
| O2 | 1.403 | 1.39 (1) | -0.02 | 0.7(2) | O2-Fe1 | 0.35 | 0.33 | -6 |
| Fe1 | 1.355 | 1.34 (1) | -0.01 | 0.5(1) | Fe1-O3 | 0.77 | 0.44 | -44 |
| O3 | 1.250 | 1.25 (1) | +0.03 | 1 | O3-Fe2 | 0.77 | 0.82 | 6 |
| Fe2 | 1.097 | 1.17 (1) | +0.03 | 1 | Fe2-O4 | 0.35 | 0.72 | 102 |
| O4 | 1.145 | 1.09 (2) | -0.02 | 1 | O4-O5 | 1.43 | 1.13 | -21 |
| O5 | 0.903 | 0.92 (1) | +0.02 | 1 | O5-Fe3 | 0.35 | 0.48 | 36 |
| Fe3 | 0.855 | 0.855 (2) | +0.00 | 1 | Fe3-O6 | 0.77 | 0.77 | 0 |

1.69(2) denotes 1.69 ± 0.02.

**Table 2** Vertical positions and occupations of atomic layers and interlayer distances obtained from fit to the vacuum-annealed surface CTR's. Vertical positions are in the unit of repeating vertical distance of 7.37 Å.

| Layer | Vertical positions and occupations | | | | Interlayer distances | | | |
|---|---|---|---|---|---|---|---|---|
| | Bulk | Nanocube | Changes | Occupations | Description | Bulk (Å) | Nanocube (Å) | Changes (%) |
| O1 | 1.597 | 1.58 (1) | -0.01 | 1.4(2) | O1-O2 | 1.43 | 1.33 | 7 |
| O2 | 1.403 | 1.40 (5) | 0.00 | 0.9(2) | O2-Fe1 | 0.35 | 0.51 | 45 |
| Fe1 | 1.355 | 1.33 (1) | -0.02 | 0.5(1) | Fe1-O3 | 0.77 | 0.96 | 23 |
| O3 | 1.250 | 1.20 (2) | -0.05 | 1 | O3-Fe2 | 0.77 | 0.61 | -22 |
| Fe2 | 1.145 | 1.12 (1) | -0.02 | 1 | Fe2-O4 | 0.35 | 0.94 | 164 |
| O4 | 1.097 | 0.99 (1) | -0.10 | 1 | O4-O5 | 1.43 | 0.35 | -76 |
| O5 | 0.903 | 0.95 (1) | 0.04 | 1 | O5-Fe3 | 0.35 | 0.69 | 96 |
| Fe3 | 0.855 | 0.85(1) | 0.00 | 1 | Fe3-O6 | 0.77 | 0.76 | -2 |

1.58(1) denotes 1.58 ± 0.01.

## 4. Discussion

Nanocrystal morphology depends on the environment during crystal formation. Theoretical calculations reported that a nanocube enclosed only with (01$\bar{1}$2) surfaces is not the most stable morphology under various pH-neutral or nearly pH-neutral natural crystal formation environments [38]. However, it was reported that during the hydrothermal synthesis, hydrolysis of DMF leads to an acidic solution [39]. The acidic environment put a limit to particle size during nanoparticle synthesis [40], which is key for monodispersive nanocube formation. Furthermore, under-coordinated surface Fe atoms will lead more acidic point of zero charge for hematite surfaces [41], which means under-coordinated Fe is favorable under the acidic environment. Hence, acidic hydrothermal synthesis conditions could generate more under-coordinated Fe atoms at the nanocube surface, which forms the half-Fe missing termination eventually. From the literature, iron oxide tends to accommodate defect through cation interstitial or vacancy and keep hexagonal oxygen layers intact [42, 43]. The half-Fe missing layer model is in line with the tendency of cation vacancy formation.

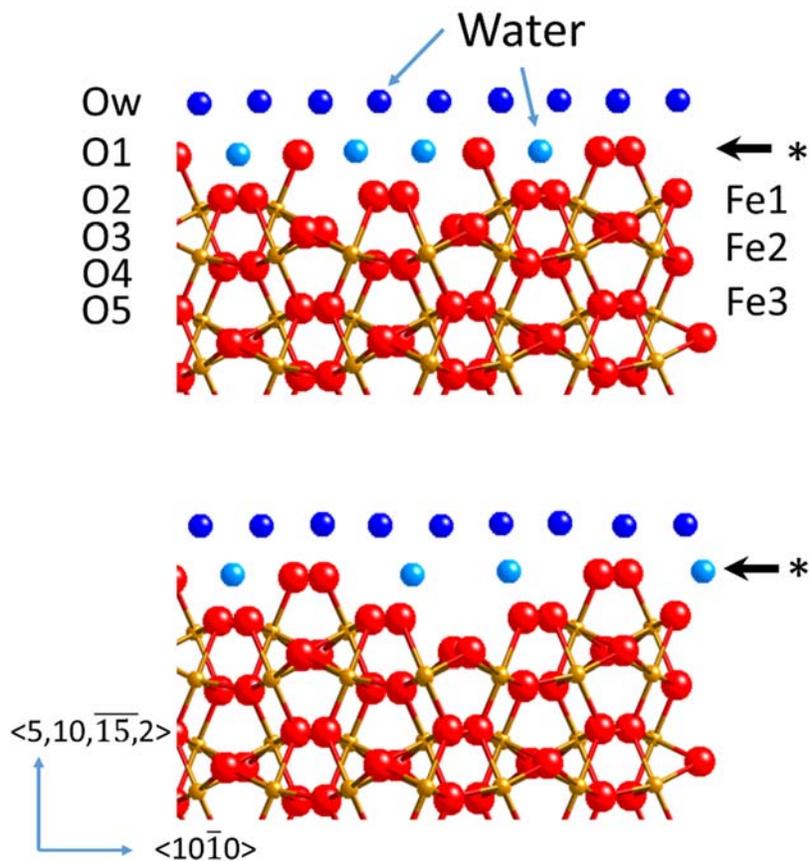

**Figure 5** Two possible (2x1) arrangements derivable from half-Fe missing model. Regular and light blue balls represent oxygen in water molecule. Light blue balls remain on the surface after annealing. The marks * indicate topmost layers in the annealed surfaces. Labelings of layers (Ow, O1, etc) are the same as those of Table 1 and 2.

It is worth noticing that the half-Fe missing model can lead to a (2x1) reconstructed surface if the missing Fe-sites are in six-fold coordination and ordered in a 2x1 fashion shown in Fig. 5. The surface can be seen as a combination of previously reported termination-E [32, 33] and oxygen layer added termination –C [31, 34]. A surface (2x1) reconstruction was observed after a single crystal ($01\bar{1}2$) surface was annealed at a high temperature under vacuum [44-47]. A recent atomic force microscope (AFM) study of vacuum annealed (2x1) surface proposed a model with ordered vacancies in the top oxygen layer [44]. However, hydrothermal acidic environment during nanocube formation is quite different from UHV annealing condition and defect accommodation mechanism could differ under the two conditions. It should be mentioned though that increasing concentrations of bulk O vacancies in sub-30 nm particles of hematite has been reported recently [48]. Deducing surface vacancy is possible through a fit of the model with occupation of atoms to CTR measurement from uniform-shape monodispersive single crystalline nanocubes. Hence, the current approach is not applicable to a usual powder samples for detection of surface vacancies [49].

Based on the CTR model fits, it can be postulated that the nanocube surfaces under hydrothermal synthesis conditions have the half-Fe missing top layer. Then, nanocube surfaces were hydrated through post-synthesis water washing and dispersion in water before the drop-cast on silicon substrate. Considering water dissociation on active sites, the topmost Fe atoms can be surrounded by additional oxygen atoms to get sixfold coordination [50].

In the half-Fe missing model for annealed surfaces shown in Fig. 5, the top-most layers (marked by *) consist of half-layer of lattice oxygens (red balls) and another half-layer of water molecules (light blue balls). The hydrated surface can be constructed by adding water layers (blue balls in Fig. 5) on top of the annealed surfaces. The results of the fit from the hydrated surface deviate from this ideal scenario as seen from the reduced occupation of oxygen atoms in the third layer from the top (O2 layer in Table 1 and Fig. 5). A certain degree of disorder in the surface region might be responsible for the deviation. Annealing improved the ordering of the surface by restoring the oxygen (O2 layer in Table 2) occupation close to one. Occupation of the topmost surface oxygen in the annealed surface (O1 layer in Table 2) is larger than one instead of half. The occupation is defined such that a full occupation of lattice atoms in the bulk structure is defined as one. If water molecules adsorb more densely than lattice oxygen, the oxygen occupation can exceed one. Thus the portion bigger than half would come from oxygens in hydrated form, shown as light blue balls in Fig. 5, as only half-layers of oxygen can be boned to Fe atoms. The annealing temperature of 115 °C does not seem high enough to remove all the water from the nanocube surface. A higher temperature annealing makes CTR intensity undetectably low, especially in the middle of two bulk peaks, and it indicates increased surface roughness as de-hydration progresses. Since the CTR measurements along the specular direction give a projection of electron density along surface normal direction, the lateral positions of the water molecules shown in Fig.5 are speculations of a possible arrangement. Theoretical calculations based on the half-Fe missing model are required to determine a 3-dimensional surface atomic structure.

## 5. Conclusions

We measured crystal truncation rods from the hydrated and annealed surfaces of hematite nanocubes and the half-Fe missing surface models fit best to the measured CTR's. Surface structures obtained from the fit can be understood with hydroxylation (annealed) of the hydrated (as-prepared) surface. The remarkably different atomic structures of the nanocube surfaces from that of macroscopic single crystals were attributed to the acidic environment during the hydrothermal nanocube synthesis process. As far as we know, this is the first surface X-ray diffraction measurement from a nanoparticle surface. Characterization of atomic structure of nanoparticle surface by X-ray diffraction could be a valuable in-situ/operando tool to monitor the surface structures of nanoparticles under reactive conditions.

## Acknowledgments

Author would like to acknowledge Garth Wells for SEM measurements. All of the research described in this paper was performed at the Canadian Light Source, a national research facility of the University of Saskatchewan, which is supported by the Canada Foundation for Innovation (CFI), the Natural Sciences and Engineering Research Council (NSERC), the National Research Council (NRC), the Canadian Institutes of Health Research (CIHR), the Government of Saskatchewan, and the University of Saskatchewan.